\documentclass[12pt]{article}
\usepackage{amssymb}
\begin{document}

%************************** Text Begins here ******************************

%  Greek letters

\def\a{\alpha}
\def\b{\beta}
\def\d{\delta}
\def\e{\epsilon}
\def\g{\gamma}
\def\h{\mathfrak{h}}
\def\k{\kappa}
\def\l{\lambda}
\def\o{\omega}
\def\p{\wp}
\def\r{\rho}
\def\t{\tau}
\def\s{\sigma}
\def\z{\zeta}
\def\x{\xi}
\def\V={{{\bf\rm{V}}}}
 \def\A{{\cal{A}}}
 \def\B{{\cal{B}}}
 \def\C{{\cal{C}}}
 \def\D{{\cal{D}}}
\def\G{\Gamma}
\def\K{{\cal{K}}}
\def\O{\Omega}
\def\R{\bar{R}}
\def\T{{\cal{T}}}
\def\L{\Lambda}
\def\f{E_{\tau,\eta}(sl_2)}
\def\E{E_{\tau,\eta}(sl_n)}
\def\Zb{\mathbb{Z}}
\def\Cb{\mathbb{C}}

\def\R{\overline{R}}
% Shorthands for \begin{equation} and the like

\def\beq{\begin{equation}}
\def\eeq{\end{equation}}
\def\bea{\begin{eqnarray}}
\def\eea{\end{eqnarray}}
\def\ba{\begin{array}}
\def\ea{\end{array}}
\def\no{\nonumber}
\def\le{\langle}
\def\re{\rangle}
\def\lt{\left}
\def\rt{\right}

\newtheorem{Theorem}{Theorem}
\newtheorem{Definition}{Definition}
\newtheorem{Proposition}{Proposition}
\newtheorem{Lemma}{Lemma}
\newtheorem{Corollary}{Corollary}
\newcommand{\proof}[1]{{\bf Proof. }
        #1\begin{flushright}$\Box$\end{flushright}}

\renewcommand{\thefootnote}{\fnsymbol{footnote}}
 \setcounter{footnote}{0}

\newfont{\elevenmib}{cmmib10 scaled\magstep1}
\newcommand{\preprint}{
   \begin{flushleft}
     %\elevenmib Yukawa\, Institute\, Kyoto\\
   \end{flushleft}\vspace{-1.3cm}
   \begin{flushright}\normalsize
  % \sf  YITP-03-53\\
   %  {\tt hep-th/yymmnnn} \\
   %July 2006
   \end{flushright}}
\newcommand{\Title}[1]{{\baselineskip=26pt
   \begin{center} \Large \bf #1 \\ \ \\ \end{center}}}
\newcommand{\Author}{\begin{center}
   \large \bf

Kun Hao${}^{a}$,~Junpeng Cao${}^{b,c}$,~Tao Yang${}^{a}$\footnote{Corresponding author: yangt@nwu.edu.cn}~and~ Wen-Li Yang${}^{a,d}$\footnote{Corresponding author: wlyang@nwu.edu.cn}

\end{center}}
\newcommand{\Address}{\begin{center}

${}^a$ Institute of Modern Physics, Northwest University,
       Xian 710069, P.R. China\\
${}^b$Beijing National Laboratory for Condensed Matter
           Physics, Institute of Physics, Chinese Academy of Sciences, Beijing
           100190, China\\
${}^c$Collaborative Innovation Center of Quantum Matter, Beijing,
     China\\
${}^d$Beijing Center for Mathematics and Information Interdisciplinary Sciences, Beijing, 100048,  China

   \end{center}}
\newcommand{\Accepted}[1]{\begin{center}
   {\large \sf #1}\\ \vspace{1mm}{\small \sf Accepted for Publication}
   \end{center}}

\preprint
\thispagestyle{empty}
\bigskip\bigskip\bigskip
\Title{Exact solution of the XXX Gaudin model with the generic open boundaries
} \Author

\Address
\vspace{1cm}

\begin{abstract}
The XXX Gaudin model with generic integrable boundaries specified by the most general
non-diagonal $K$-matrices is studied by the off-diagonal Bethe ansatz method.
The eigenvalues of the associated Gaudin operators and the corresponding Bethe ansatz equations are obtained.

\vspace{1truecm} \noindent {\it PACS:} 03.65.Fd; 04.20.Jb; 05.30.-d; 75.10.Jm

\end{abstract}

\newpage

\setcounter{footnote}{0}
\renewcommand{\thefootnote}{\arabic{footnote}}

\baselineskip=20pt

\section{Introduction}
\label{intro} \setcounter{equation}{0}

The Gaudin models introduced in 1976 by M. Gaudin \cite{Gau76}, describe completely integrable classical and quantum long-range interacting spin chains. It has
played a distinguished role in many areas of modern physics \cite{Garraway11,Kulish12}, such as in establishing the integrability of the
Seiberg-Witten thoery \cite{Sei94, Bra99}, constructing the integral representations of the solutions to the Knizhnik-Zamolodchikov (KZ) equation \cite{KZ,Hik95,H.M93,B. Feign94} and  using as a testing ground for ideas such as the functional Bethe ansatz and general procedure of separation of variables \cite{Skl87,Skl96,Skl99}.

The XXX Gaudin model was studied by Sklyanin \cite{Skl87} and Jur\v{c}o \cite{Jour1989} from the point of view of the quantum inverse scattering method and the corresponding eigenvalues and the eigenfunctions of the model were obtained. In 1992 Hikami used the transfer matrix of the periodic chain to construct a variety of conserved quantities  (or Gaudin operators) for the Gaudin model \cite{K. Hikami92}. Twisted rings can be cut to open chains and loops include two reflections at the boundaries. The possibility to include such reflections in integrable theory was founded and systematically investigated by Sklyanin \cite{Skl88}.
In particular, Hikami \cite{Hik95} constructed the the eigenvalue and eigenfunctions for the Gaudin model with special open boundary condition which is associated with diagonal K-matrices (which are the solutions to the reflection equations \cite{Cherednik84}). This method was then used to get the exact solution of the related BCS model \cite{Ami01}, the integrable XXZ Gaudin model with some restricted non-diagonal $K$-mdtrices \cite{Yang04a}, and the boundary integrable Gaudin models associated with higher rank algebras \cite{Yan04,Yan05}. On the other hand, to obtain correlation functions, we need to calculate the scalar product of Bethe states \cite{Kor93}. In 2002, Zhou et al. \cite{Zho02}, by using the relationship between periodic Gaudin model and spin chain, constructed the determinant representation of scalar product of Bethe state with periodic Gaudin model.  Recently,   the determinant representation of scalar product of Bethe state for the XXZ Gaudin model with some special non-diagonal boundaries were  successfully constructed \cite{Kun Hao2012}.

The relationship between Gaudin models and BCS models lies in the fact that the exact solution of BCS model was constructed long ago by Richardson and Sherman by a procedure close in spirit to the coordinate Bethe ansatz (BA)\cite{Rich63}. The knowledge of the exact eigenstates and eigenvalues of the BCS model has been crucial to establish physically relevant observables, in particular it was used to study small metallic grains \cite{Ami01,Falci98}; the picture was merged in the scenario of QISM in \cite{Amico01}. The integrability of the model has been proved \cite{Skl87} to be deeply related to the integrability of the isotropic Gaudin magnet \cite{Gau76}: the BCS model can be expressed as a certain combination of its integrals of motion, which contain Gaudin Hamiltonians (or  Gaudin operators).
Connection with conformal field theory and WZNW models in field theory have been deeply investigated \cite{Sierra00} based on the relation between solution of KZ equations and Gaudin model found in \cite{H.M93,Reshetikhin95}. The class of pairing Hamiltonians was generalized by investigating the quasi-classical expansion of the disordered twisted six vertex model $R$-matrix. Many properties of metallic grains in a normal state can be described by the orthodox model \cite{DOTS,AVERIN}.

It has been well known for many years that there exists a quite usual class of integrable models \cite{Cao03,bax1,Tak79,Yun95,Nep04}, which do not possess $U(1)$-symmetry and thus make the conventional Bethe ansatz methods \cite{Skl88,bax1,Tak79,Bet31,Skl78,Alc87} almost inapplicable. Recently, a systematic method, namely, the off-diagonal Bethe ansatz (ODBA) method \cite{Cao1} was proposed to approach the exact solutions of generic integrable models either with or without $U(1)$ symmetry. Several long-standing models were then solved \cite{Cao1,Li14,Zha13,Hao14} by the ODBA method. Some other interesting methods for approaching such kind of problems are
the q-Onsager algebra method \cite{Bas07} and  the Sklyanin's separation of variables (SOV) method \cite{Skl92} which has been recently applied to some integrable models  related to $su(2)$ algebra
\cite{Nic12,Fad14,Kit14,Fra08}.

In this paper we shall study the XXX Gaudin model with the most general non-diagonal integrable boundary condition which corresponds to arbitrary unparallel boundary fields via the ODBA method. The paper is organized as follows. The next section serves as an introduction of some basic ingredients of the transfer matrix of the inhomogeneous spin-$\frac{1}{2}$ open XXX chain  and the associated Gaudin type model. In Section 3 the  XXX Gaudin operators associated with the most general non-diagonal integrable boundary condition are constructed. After a briefly reviewing the
ODBA method with  our parameterization,  we obtain the eigenvalues of the resulting Gaudin operators and the corresponding Bethe ansatz equations (BAEs) in Section 4. We summarize our results and give some discussions in Section 5.

\section{Inhomogeneous open XXX  chain}
\setcounter{equation}{0}

The rational six-vertex model R-matrix $R(u)\in {\rm End}({\rm\bf V}\otimes {\rm\bf V})$
is given by
\bea
R(u)=\lt(\begin{array}{llll}u+\eta&&&\\&u&\eta&\\
&\eta&u&\\&&&u+\eta\end{array}\rt),
\label{r-matrix}\eea
with $u$ being the spectral parameter and $\eta$ being the crossing parameter.
The $R$-matrix satisfies the quantum Yang-Baxter equation (QYBE)
\bea
R_{12}(u_1-u_2)R_{13}(u_1-u_3)R_{23}(u_2-u_3)
=R_{23}(u_2-u_3)R_{13}(u_1-u_3)R_{12}(u_1-u_2),\label{QYB}\eea
and the properties,
\bea &&\hspace{-1.5cm}\mbox{Initial
condition}:\,R_{12}(0)= \eta P_{12},\label{Int-R}\\
&&\hspace{-1.5cm}\mbox{Unitarity
relation}:\,R_{12}(u)R_{21}(-u)= -\xi(u)\,{\rm id},
\quad \xi(u)=(u+\eta)(u-\eta),\label{Unitarity}\\
&&\hspace{-1.5cm}\mbox{Crossing
relation}:\,R_{12}(u)=V_1R_{12}^{t_2}(-u-\eta)V_1,\quad
V=-i\s^y,
\label{crosing-unitarity}\\
&&\hspace{-1.5cm}\mbox{Quasi-classical property}:R(u)|_{\eta\rightarrow0}=u\times {\rm id},\\
&&\hspace{-1.5cm}\mbox{Antisymmetry}:\,R_{12}(-\eta)=-\eta(1-P)=-2\eta P^{(-)}.\label{Ant}
\eea
Here $R_{21}(u)=P_{12}R_{12}(u)P_{12}$ with $P_{12}$ being
the usual permutation operator and the superscript $t_i$ denotes transposition
in the $i$-th space. Here and below we adopt the standard
notations: for any matrix $A\in {\rm End}({\rm\bf V})$, $A_j$ is an
embedding operator in the tensor space ${\rm\bf V}\otimes
{\rm\bf V}\otimes\cdots$, which acts as $A$ on the $j$-th space and as
identity on the other factor spaces; $R_{ij}(u)$ is an embedding
operator of R-matrix in the tensor space, which acts as identity
on the factor spaces except for the $i$-th and $j$-th ones.

Let us  introduce two ``row-to-row" monodromy matrices $T(u)$ and $\hat{T}(u)$,
which are $2\times 2$ matrix with elements being
operators acting on ${\rm\bf V}^{\otimes N}$,
\bea
T_0(u)&=&R_{0N}(u-\theta_N)R_{0\,N-1}(u-\theta_{N-1})\cdots
R_{01}(u-\theta_1),\label{Mon-V-1}\\
\hat{T}_0(u)&=&R_{01}(u+\theta_1)R_{02}(u+\theta_{2})\cdots
R_{0N}(u+\theta_N).\label{Mon-V-2}
\eea
Integrable open chain can be constructed as follows \cite{Skl88}.
Let us introduce a pair of $K$-matrices $K^-(u)$ and $K^+(u)$. The former satisfies the reflection equation (RE)
 \bea &&R_{12}(u_1-u_2)K^-_1(u_1)R_{21}(u_1+u_2)K^-_2(u_2)\no\\
 &&~~~~~~=
K^-_2(u_2)R_{12}(u_1+u_2)K^-_1(u_1)R_{21}(u_1-u_2),\label{RE-V}\eea
and the latter satisfies the dual RE
\bea
&&R_{12}(u_2-u_1)K^+_1(u_1)R_{21}(-u_1-u_2-2)K^+_2(u_2)\no\\
&&~~~~~~= K^+_2(u_2)R_{12}(-u_1-u_2-2)K^+_1(u_1)R_{21}(u_2-u_1).
\label{DRE-V}\eea
For open spin-chains, other than the standard
``row-to-row" monodromy matrix $T(u)$ (\ref{Mon-V-1}), one needs to
consider  the
 double-row monodromy matrix $\mathbb{T}(u)$
\bea
  \mathbb{T}(u)=T(u)K^-(u)\hat{T}(u).
  \label{Mon-V-0}
\eea
Then the double-row transfer matrix of the XXX chain with open
boundary is given by
\bea
\t(u)=tr(K^+(u)\mathbb{T}(u)).\label{trans}
\eea
With the help of QYBE (\ref{QYB}) and (dual) REs
(\ref{RE-V}) and (\ref{DRE-V}),
one can prove \cite{Skl88} that the transfer matrices with
different spectral parameters commute with each other:
\bea
[\t(\mu),\t(\nu)]=0.
\eea
Then $\tau(u)$ is the generating function of the conserved quantities of
the corresponding system, which ensures the integrability of the open XXX chain.

\section{XXX Gaudin model with generic boundaries}
\setcounter{equation}{0}

In this paper, we consider the most general $K$-matrix  which is most general solution to
the RE (\ref{RE-V}) \cite{Veg93}
\bea
K^-(u)=\xi+u\vec{h}_1\cdot\vec{\sigma},\quad \vec{h}_1^2=1,\label{k-}
\eea
where $\vec{\sigma}=(\sigma_x,\sigma_y,\sigma_z)$ and $\sigma_{\alpha}$ are the Pauli matrices, and $\vec{h}_1$ is a normalized three-dimensional
vector which are related to one boundary field. At the same time we consider the most general  dual $K$-matrix $K^+(u)$
\bea
K^+(u)=\bar{\xi}+(u+\eta)\vec{h}_2\cdot\vec{\sigma},\quad \vec{h}_2^2=1.\label{k+}
\eea
We remark that the  transfer matrix (\ref{trans}) indeed does depend six free boundary parameters, i.e,
$\xi$, $\vec{h}_1$, $\bar{\xi}$ and $\vec{h}_2$ due to the fact we have normalized the two three-dimensional vectors $\vec{h}_i$ to unity.

Following \cite{Skl87, K. Hikami92, Yang04a}, one can construct the associated Gaudin operators by expanding the transfer matrix $\t(\theta_j)$ around
small $\eta$. In order that the resulting Gaudin operators form a commuting family, which ensures the integrability of the corresponding Gaudin model \cite{Gau76},
one need to give some constraints of the $\eta$-dependence of the boundary parameters. Without losing the generality,   one can assume that the boundary
parameters $\xi$ and $\vec{h}_1$ do not depend on the crossing parameter $\eta$ but $\bar{\xi}$ and $\vec{h}_2$ do. Let us expand $\bar{\xi}$ and $\vec{h}_2$
with respect to $\eta$ as follows \footnote{There also exists another choice of the expansions $\vec{h}_2=-\vec{h}_1+\vec{h}_2^{(1)}\eta+\frac{\vec{h}_2^{(2)}}{2!}\eta^2+o(\eta^3)$
and $\bar{\xi}=\xi+\xi^{(1)}\eta+o(\eta^2)$ such that the identity (\ref{zero-term}) is fulfilled. One can deal with the case by the similar method as that in the following part of this paper.}:
\bea
\vec{h}_2(\eta)&=&\vec{h}_1+\vec{h}_2^{(1)}\eta+\frac{\vec{h}_2^{(2)}}{2!}\eta^2+o(\eta^3),\label{h2}\\
\bar{\xi}(\eta)&=&-\xi+\xi^{(1)}\eta+o(\eta^2).\label{Parameter-1}
\eea
Moreover, the normalized condition (\ref{k+}) of the vector $\vec{h_2}$ implies that
\bea
\vec{h}_1\cdot \vec{h}_2^{(1)}=0,\quad \vec{h}_1\cdot \vec{h}_2^{(2)}=-\vec{h}_2^{(1)}\cdot \vec{h}_2^{(1)},\label{Orthono}
\eea which means that the vector $\vec{h}_2^{(1)}$ (or $\vec{h}_2^{(2)}$) has only two degrees of freedom.
The above expansion leads to the following identity
\bea
\lim_{\eta\rightarrow0}\{K^+(u)K^-(u)\}=\lim_{\eta\rightarrow0}\{K^+(u)\}K^-(u)=f(u)\times {\rm id}.\label{zero-term}
\eea

The corresponding  XXX Gaudin operators can be obtained by expanding the double-row transfer
matrix $\tau(u)$ at the point $u = \theta_j$ around $\eta=0$:
\bea
\t(\theta_j)=\eta(\t(\theta_j)^{(0)}+\eta H_j+\ldots).\label{Expandsion-eta}
\label{exp-t}\eea
Direct calculation shows that the coefficient of the leading term in the above expansion is
\bea
\t(\theta_j)^{(0)}&=&\lim_{\eta\rightarrow 0} tr \lt\{
\prod_{i=1,i\neq j}^{N}(\theta_j-\theta_i)\prod_{i=1}^{N}(\theta_j+\theta_i)K^+_0(\theta_j)P_{0j}K^-_0(\theta_j)\rt\}\no\\
&=&\prod_{i=1,i\neq j}^{N}(\theta_j-\theta_i)\prod_{i=1}^{N}(\theta_j+\theta_i)\times\lim_{\eta\rightarrow0}\{K^+_j(\theta_j)K^-_j(\theta_j)\}\no\\
&\stackrel{(\ref{zero-term})}{=}&\prod_{i=1,i\neq j}^{N}(\theta_j-\theta_i)\prod_{i=1}^{N}(\theta_j+\theta_i)\,(\theta_j^2-\xi^2)\times{\rm id}.\label{Leading-term}
\eea
Then the XXX Gaudin operators $\{H_j|j=1,\cdots,N\}$ with generic open boundaries  specified by the boundary $K$-matrices in (\ref{k-}) and (\ref{k+})
defined by (\ref{Expandsion-eta}) are given by
\bea
H_j&=&\lt.\frac{\partial(\t(\theta_j)/\eta)}{\partial\eta}\rt|_{\eta=0}\no\\
&=&\prod_{i=1,i\neq j}^{N}(\theta_j-\theta_i)\prod_{i=1}^{N}(\theta_j+\theta_i)\lt[\xi\xi^{(1)}+\theta_j-\frac{\xi^2}{\theta_j}
+i\theta_j^2(\vec{h}_1\times\vec{h}_2^{(1)})\cdot\vec{\sigma}_j\rt.\no\\
&&+\theta_j(\xi^{(1)}\vec{h}_1+\xi\vec{h}_2^{(1)})\cdot\vec{\sigma}_j
+\sum_{k=1,k\neq j}^N\frac{\theta_j^2-\xi^2}{2(\theta_j-\theta_k)}
(\sigma^x_j\sigma^x_k+\sigma^y_j\sigma^y_k+\sigma^z_j\sigma^z_k+{\rm id})\no\\
&&\lt.+\sum_{k=1,k\neq j}^N\frac{(\xi+\theta_j\vec{h}_1\cdot\vec{\sigma}_j)}{2(\theta_j+\theta_k)}(\sigma^x_j\sigma^x_k+
\sigma^y_j\sigma^y_k+\sigma^z_j\sigma^z_k+{\rm id})(-\xi+\theta_j\vec{h}_1\cdot\vec{\sigma}_j)\rt]
.\label{off-gaudin-H}
\eea
In the above derivation, we have used the following relations
\bea
K^-_j(\theta_j)\,\frac{\partial(K^+_j(\theta_j))}{\partial\eta}|_{\eta=0}&=&
(\xi+\theta_j\vec{h}_1\cdot\vec{\sigma}_j)(\xi^{(1)}+(\vec{h}_1+\theta_j\vec{h}_2^{(1)})\cdot\vec{\sigma}_j) ,\\[2pt]
(\vec{a}\cdot\vec{\sigma})(\vec{b}\cdot\vec{\sigma})
&=&\vec{a}\cdot\vec{b}+i\vec{\sigma}\cdot(\vec{a}\times\vec{b}),
\eea
where $\vec{a}\times\vec{b}$ denotes the cross product of two three-dimensional vectors.
It follows from the explicit expressions of the Gaudin operators $\{H_j\}$ given by (\ref{off-gaudin-H})
that in addition to the  inhomogeneous parameters $\{\theta_j\}$, these operators indeed depend on  the  six free parameters $\{\xi,\xi^{(1)},\vec{h}_1,\vec{h}_2^{(1)}\}$. The parameters $\{\xi^{(1)},\vec{h}_2^{(1)}\}$ are given by (\ref{h2}) and (\ref{Parameter-1}).

The fact that the leading term of (\ref{Expandsion-eta}) is proportion to the identity operator (see (\ref{Leading-term}))  and  the commutativity of the transfer matrices (\ref{trans}) with different spectrum parameters for a generic $\eta$ implies
\bea
[H_i,H_j]=0,\qquad i,j=1,2,\cdots,N.
\eea
Thus the Gaudin model defined by (\ref{off-gaudin-H}) is integrable.
The main aim of this paper is to find the eigenvalues  of the Gaudin operators and the corresponding  BAEs.

\section{Eigenvalues and the Bethe ansatz equations}
\setcounter{equation}{0}
The fact that the Gaudin operators $\{H_j\}$ (\ref{off-gaudin-H}) can be expressed in terms of the transfer matrix of the inhomogeneous
spin-$\frac{1}{2}$ XXX open chain enables us to extract eigenvalues of the Gaudin operators and the associated Bethe ansatz equations from those
of the XXX open chain.

\subsection{Off-diagonal Bethe ansatz solution of the  XXX open chain}

Thanks to the works \cite{Cao1}, the transfer matrix (\ref{trans}) of the open XXX chain with generic open boundaries  specified by the boundary $K$-matrices in (\ref{k-}) and (\ref{k+}) can be exactly diagonalized by off-diagonal Bethe ansatz method as follows. Let $\Lambda(u)$ be an eigenvalue of the transfer matrix
$\tau(u)$ given by (\ref{trans}).  Following the same procedure as in \cite{Cao1}, we can show that  $\Lambda(u)$ satisfies the following functional relations
\bea
\Lambda(\theta_j)\Lambda(\theta_j-\eta)&=&\frac{4(\theta_j+\eta)(\theta_j-\eta)(\xi^2-\theta^2_j)(\theta^2_j-\bar{\xi}^2)}{(2\theta_j-\eta)(2\theta_j+\eta)}\no\\[2pt]
&&\times\prod^N_{i=1}(\theta_j+\theta_i+\eta)(\theta_j-\theta_i+\eta)\prod^N_{i=1}(\theta_j+\theta_i-\eta)(\theta_j-\theta_i-\eta),\no\\[2pt]
&&\quad j=1,\ldots,N,\label{Main}
\eea
and the following properties
\bea
&&\hspace{-1.5cm}\mbox{Crossing
symmetry}:\,\quad\L(-u-\eta)=\L(u),\label{Eig-Cro}\\
&&\hspace{-1.5cm}\mbox{Initial
condition}:\,\L(0)=2\xi\,\bar{\xi}\prod_{j=1}^N(\eta-\theta_j)(\eta+\theta_j)=\L(-\eta),\label{Eig-In}\\
&&\hspace{-1.5cm}\mbox{Asymptotic behavior}:\,
\L(u)= 2\vec{h}_1\cdot\vec{h}_2u^{2N+2}+\ldots,\quad {\rm for}\, u\rightarrow\infty,\label{Eig-Asy} \\
&&\hspace{-1.5cm}\mbox{Analyticity}:\L(u) \mbox{, as a function of $u$, is a polynomial of degree $2N+2$}.\label{Eigen-Anal}
\eea
The above conditions (\ref{Main})-(\ref{Eigen-Anal}) completely determine the eigenvalue  $\L(u)$, which allows one to
construct the off-diagonal Bethe ansatz solution of the XXX open chain. Namely, $\L(u)$ can be given in terms of certain inhomogeneous
$T-Q$ relation \cite{Cao1}.  Moreover, it is known that the $T-Q$ relation can be written in many different types which corresponds to
the different parameterizations of the solutions of (\ref{Main})-(\ref{Eigen-Anal}), and each of them gives the
complete set of eigenvalues of the transfer matrix.
Here we present a simple $T-Q$ relation for $\Lambda(u)$, which corresponds to the $M=0$ type in \cite{Cao1}, namely we can construct the solution of (\ref{Main})-(\ref{Eigen-Anal}) by the following ansatz
\bea
\Lambda(u)=\bar{a}(u)\frac{Q(u-\eta)}{Q(u)}+
\bar{d}(u)\frac{Q(u+\eta)}{Q(u)}+c\,u(u+\eta)\frac{F(u)}{Q(u)},
\label{T-Q}
\eea
with the function $Q(u)$ is parameterized by $N$ parameters $\{\lambda_j|j=1,\ldots,N\}$
\bea
 Q(u)=\prod_{j=1}^{N}(u-\lambda_j)(u+\lambda_j+\eta)=Q(-u-\eta),\no
\eea
and the functions $\bar{a}(u)$, $\bar{d}(u)$, $F(u)$ and the constant $c$ are given by respectively
\bea
&\bar{a}(u)&=\frac{2u+2\eta}{2u+\eta}(u+\xi)(u+ \bar{\xi})\prod_{j=1}^{N}(u+\theta_j+\eta)(u-\theta_j+\eta),\label{a-function}\\
&\bar{d}(u)&=\frac{2u}{2u+\eta}(u-\xi+\eta)((u+\eta)-\bar{\xi})
\prod_{j=1}^{N}(u+\theta_j)(u-\theta_j)\no\\
&&\quad\quad\quad =\bar{a}(-u-\eta),\label{d-function}\\
&F(u)&=\prod_{j=1}^N(u+\theta_j)(u-\theta_j)(u+\theta_j+\eta)(u-\theta_j+\eta),\\
&c&=2(\vec{h}_1\cdot\vec{h}_2-1).
\eea
The analyticity of $\Lambda(u)$ requires the apparent simple poles $u=\lambda_j,\,j=1\ldots,N$ are not real poles, or the residues at these points must vanish, which lead to the following BAEs
\bea
\bar{a}(\lambda_j)Q(\lambda_j-\eta)+
\bar{d}(\lambda_j)Q(\lambda_j+\eta)+c\,\lambda_j(\lambda_j+\eta)F(\lambda_j)
=0,\qquad
j=1,\ldots,N.\label{BAE-M}
\eea
It is easy to check that the $T-Q$ relation (\ref{T-Q}) does satisfies the relations (\ref{Main})-(\ref{Eigen-Anal}) if the $N$ parameters $\{\lambda_j|j=1\ldots,N\}$ satisfy the BAEs (\ref{BAE-M}).

\subsection{Eigenvalues of the Gaudin operators}
The relation (\ref{exp-t}) implies that the Gaudin operators  $\{H_j\}$ given by (\ref{off-gaudin-H}) can be expressed in terms of the expansion of transfer matrix of the inhomogeneous spin-$\frac{1}{2}$ XXX open chain around small $\eta$ and that one can extract eigenvalues of the Gaudin operators and the associated Bethe ansatz equations from those
of the XXX open chain.

For this purpose, let us evaluate the function (\ref{T-Q}) at points $\theta_j$ for small $\eta$
\bea
\Lambda(\theta_j)&=&\frac{2\theta_j+2\eta}{2\theta_j+\eta}
(\theta_j+\xi)(\theta_j+\bar{\xi})\prod^N_{i=1}
(\theta_j+\theta_i+\eta)(\theta_j-\theta_i+\eta)\no\\[2pt]
&&\times\frac{\prod^{N}_{i=1}(\theta_j-\lambda_i-\eta)(\theta_j+\lambda_i)}
{\prod^{N}_{i=1}(\theta_j-\lambda_i)(\theta_j+\lambda_i+\eta)}\no\\[2pt]
&=&\eta\lt\{(2\theta_j+2\eta)
(\theta_j+\xi)(\theta_j+\bar{\xi})\prod^N_{i\neq j}
(\theta_j+\theta_i+\eta)(\theta_j-\theta_i+\eta)\rt.\no\\[2pt]
&&\times\quad \lt.\frac{\prod^{N}_{i=1}(\theta_j-\lambda_i-\eta)(\theta_j+\lambda_i)}
{\prod^{N}_{i=1}(\theta_j-\lambda_i)(\theta_j+\lambda_i+\eta)}+\eta\,E_j+\ldots\rt\}
,\qquad j=1,\cdots,N.\eea
With the help of the relation (\ref{exp-t}) and the above expansion, the eigenvalue $E_j$ of the Gaudin operators $H_j$ are then given by
\bea
E_j&=&(\theta_j^2-\xi^2)(2\theta_j)\hspace{-0.20truecm}
\prod_{i=1,i\neq j}^{N}(\theta_j+\theta_i)(\theta_j-\theta_i)
\lt[\frac{\xi^{(1)}}{\theta_j-\xi}
+\hspace{-0.20truecm}\sum_{i=1,i\neq j}^N(\frac{1}{\theta_j+\theta_i}+\frac{1}{\theta_j-\theta_i})\rt.\no\\[2pt]
&&\lt.
+{1\over\theta_j}+\sum_{i=1}^{N}(\frac{-1}{\theta_j-\lambda_i}+\frac{-1}{\theta_j+\lambda_i})\rt],\label{Eignvalue-Ej}
\eea
where the parameter $\xi^{(1)}$ is given by (\ref{Parameter-1}) and the $N$ parameters $\{\l_j|j=1,\ldots,N\}$
need to satisfy the associated  BAEs:
\bea
\frac{1-\xi^{(1)}}{\lambda_j-\xi}&+&\frac{1-\xi^{(1)}}{\lambda_j+\xi}
-\sum^N_{i=1}(\frac{1}{\lambda_j+\theta_i}
+\frac{1}{\lambda_j-\theta_i})+\sum^{N}_{i=1,i\neq j}(\frac{2}{\lambda_j-\lambda_i}+\frac{2}{\lambda_j+\lambda_i})\no\\[2pt]
&=&\frac{\vec{h}^{(1)}_2\cdot{\vec{h}_2^{(1)}}\lambda_j}
{2(\lambda_j^2-\xi^2)}
\frac{\prod_{l=1}^N(\lambda_j+\theta_l)(\lambda_j-\theta_l)}
{\prod_{i=1,i\neq j}^{N}(\lambda_j-\lambda_i)(\lambda_j+\lambda_i)}
,\qquad j=1,\cdots,N.\label{BAE-G}\eea
In the derivation of the above BAEs, we have used the relation (\ref{Orthono}).

Some remarks are in order. In addition to the inhomogeneous parameters $\{\theta_j\}$
, the associated Gaudin operators $\{H_j\}$ given by (\ref{off-gaudin-H}), depend on  the following six free parameters
$\{\xi,\xi^{(1)},\vec{h}_1,\vec{h}_2^{(1)}\}$. These six parameters all enter the associated BAEs (\ref{BAE-G}). When two vectors are parallel
to each other, i.e. $\vec{h}_2=\vec{h}_1$, the resulting Gaudin operators and their eigenvalues and the associated BAEs
recover those \cite{K. Hikami92} obtained by algebraic Bethe ansatz method.

\section{Conclusions}
\label{EMT} \setcounter{equation}{0}

In this paper we have studied the XXX Gaudin model with the generic open boundaries specified by the most general
$K$-matrices (\ref{k-}) and (\ref{k+}) respectively. The associated Gaudin operators are given by (\ref{off-gaudin-H}).
Besides the inhomogeneous parameters $\{\theta_j\}$, these Gaudin operators have six free parameters $\{\xi,\xi^{(1)},\vec{h}_1,\vec{h}_2^{(1)}\}$,
which lead to six-parameter $\{\xi,\xi^{(1)},\vec{h}_1,\vec{h}_2^{(1)}\}$ generalizations of those in \cite{Ami01,Zho02} and five-parameter $\{\xi^{(1)},\vec{h}_1,\vec{h}_2^{(1)}\}$ (the case of non-vanishing $\vec{h}^{(1)}_2$) generalizations of those in \cite{Hik95, Lor02, Yang04a, Kun Hao2012}. The eigenvalues
of the operators are given by (\ref{Eignvalue-Ej}) with help of  the off-diagonal Bethe ansatz method. The corresponding  BAEs are given by (\ref{BAE-G}).

\section*{Acknowledgments}

The financial support from  the National Natural Science Foundation
of China (Grant Nos. 11174335, 11031005, 11375141, 11374334), the
National Program for Basic Research of MOST (973 project under grant
No.2011CB921700), the State Education Ministry of China (Grant No.
20116101110017) and BCMIIS are gratefully acknowledged.


\begin{thebibliography}{99}

\bibitem{Gau76} M. Gaudin, {\it J. Phys.} (Paris) {\bf 37} (1976), 1087.

\bibitem{Garraway11} B. M. Garraway, {\it Phil.Trans. R. Soc.} A {\bf 369} (2011), 1137.
\bibitem{Kulish12} N. M. Bogoliubov and P. P. Kulish, {\it Zap. Nauchn. Sem. S.-Peterburg. Otdel. Mat. Inst. Steklov.}
(POMI) (Russian) {\bf 398} (2012), 26.

\bibitem{Sei94} N. Seiberg and E. Witten, Nucl. Phys. {\bf B 426} (1994), 19.

\bibitem{Bra99} H. Braden, A. Marshakov, A. Mirohov and A. Morozov,
    {\em The Ruijsenaars-Schneider model in the context of
    Seiberg-Witten theory}, {\tt e-print hep-th/9902205}.

\bibitem{KZ} V.G. Knizhnik and A.B. Zamolodchikov, Nucl. Phys. {\bf B 247} (1984), 83.
\bibitem{Hik95} K. Hikami, J. Phys {\bf A 28} (1995), 4997.
\bibitem{H.M93} H.M. Babujian, {\it J. Phys.}  {\bf A 26} (1993), 6981.
\bibitem{B. Feign94} B. Feign, E. Frenkel and N. Reshetikhin, {\it Commun. Math. Phys.} {\bf 166} (1994), 27.

\bibitem{Skl87} E.K. Sklyanin, J. Sov. Math {\bf 47} (1989), 2473.

\bibitem{Skl96} E. K. Sklyanin and T. Takebe, Phys. Lett. {\bf A 219} (1996), 217.

\bibitem{Skl99} E.K. Sklyanin, Lett. Math. Phys. {\bf 47} (1999), 275.

\bibitem{Jour1989} B. Jur\v{c}o, {\it Jour. Math. Phys. }  {\bf 30} (1989), 1289.

\bibitem{K. Hikami92} K. Hikami, P.P. Kulish, M. Wadati, {\it J. Phys. Soc. Jpn.} {\bf 61} (1992), 3071.

\bibitem{Skl88} E. K. Sklyanin, {\it J. Phys.} {\bf A 21} (1988), 2375.

\bibitem{Cherednik84} I.V. Cherednik, {\it Thero. Math. Fiz.} {\bf 61} (1984), 35.

\bibitem{Ami01} L. Amico, A. Di Lorenzo and A. Osterloh, {\it
Phys. Rev. Lett.\/} {\bf 86} (2001), 5759; \\
L. Amico, A. Di Lorenzo and A. Osterloh, {\it Nucl. Phys.} {\bf B
614} (2001), 449.

\bibitem{Yang04a} W.\,-L. Yang, Y.\,-Z. Zhang and M. Gould, {\it Nucl. Phys.\/} {\bf B 698} (2004), 312.

\bibitem{Yan04} W.\,-L. Yang, R. Sasaki and Y.\,-Z. Zhang,  {\it JHEP\/}
{\bf 09}  (2004), 046.
\bibitem{Yan05} W.\,-L. Yang, Y.\,-Z. Zhang and R. Sasaki, {\it Nucl. Phys.\/} {\bf B 729}
(2005), 594.

\bibitem{Kor93} V.\,E. Korepin, N.\,M. Bogoliubov and A.\,G. Izergin,
{\it Quantum Inverse Scattering Method and correlation
Function\/}, Cambridge Univ. Press, Cambridge, (1993).

\bibitem{Zho02} H.\,Q. Zhou, J.\,R. Links, R. Mackenzie and M.\,D.
Gould, {\it Phys. Rev.\/} {\bf B 65} (2002), 060502.


\bibitem{Kun Hao2012} K. Hao,W.\,-L. Yang, H. Fan, S.\,-Y. Liu, K. Wu, Z.\,-Y. Yang and
Y.\,-Z. Zhang,  {\it Nucl. Phys. \/} {\bf B 862} (2012), 835.

\bibitem{Rich63}R.\,W. Richardson,  {\it Phys. Lett.\/} {\bf A 3} (1963), 277;\\
R.\,W. Richardson, {\it Phys. Lett.\/} {\bf B 5} (1963), 82;\\
R.\,W. Richardson, N.\, Sherman, {\it Nucl. Phys.\/} {\bf52} (1964), 221.

\bibitem{Falci98}G.\,Falci, and R. \, Fazio,{\it Phys. Rev. Lett.\/} {\bf 80} (1998), 4542.

\bibitem{Amico01}L. Amico, G.~Falci, and R. Fazio, {\it J. Phys.} {\bf A 34} (2001), 6425.

\bibitem{Sierra00}G. Sierra, Nucl.~Phys.~B {\bf 572} (2000), 517.

\bibitem{Reshetikhin95}N.~Reshetikhin and A.~Varchenko.
\newblock In {\em Geometry, Topology, and Physics}, page 293 (1995).

\bibitem{DOTS}{\it Proceedings of the International
Conference on Electron Transport in Mesoscopic Systems, ETMS'99}
 ~J. Low Temp. Phys. {\bf 118} (2000).

\bibitem{AVERIN} D. Averin and K.~K.~Likharev, in {\it Mesoscopic phenomena
 in solids}, eds. B.L. Altshuler, P. A. Lee, and R. A. Webb (Elsevier,
 New York, 1991).

\bibitem{Cao03} J. Cao, H.\,-Q. Lin, K.\,-J. Shi and Y. Wang, {\it Nucl. Phys.\/} {\bf B 663} (2003), 487.

\bibitem{bax1} R. J. Baxter, {\it Phys. Rev. Lett.} {\bf 26} (1971), 832;\\
R. J. Baxter, {\it Phys. Rev. Lett.} {\bf 26} (1971), 834; \\
R. J. Baxter, {\it Ann. Phys. (N.Y.)} {\bf 70} (1967), 323.

\bibitem{Tak79} L. A. Takhtadzhan and L. D. Faddeev, Rush. Math.
Surveys {\bf 34} (1979), 11.

\bibitem{Yun95} C.M. Yung, M.T. Batchelor, Nucl. Phys. {\bf B 446} (1995), 461.
\bibitem{Nep04} R.\,I. Nepomechie, {\it J. Phys.\/} {\bf 34}
(2001), 9993;\\
R.\,I. Nepomechie,  {\it Nucl. Phys.} {\bf B 622} (2002), 615;\\
R.\,I. Nepomechie, {\it J. Stat. Phys.\/} {\bf 111} (2003), 1363;\\
R.\,I. Nepomechie, {\it J. Phys.\/} {\bf A 37} (2004), 433.

\bibitem{Bet31} H. Bethe, {\it Z. Phys.} {\bf 71} (1931), 205.

\bibitem{Skl78} E. K. Sklyanin and L. D. Faddeev, Sov. Phys. Dokl. {\bf 23} (1978), 902.

\bibitem{Alc87} F.\,C. Alcaraz, M.\,N. Barber, M.\,T. Batchelor,
R. \,J. Baxter and G.\,R.\,W. Quispel, {\it J. Phys.} {\bf A 20} (1987), 6397.

\bibitem{Cao1} J. Cao, W.-L. Yang, K. Shi and Y. Wang, {\it  Phys. Rev. Lett.} {\bf 111} (2013), 137201; \\
J. Cao, W.-L. Yang, K. Shi and Y. Wang, {\it Nucl. Phys. }{\bf B 875} (2013), 152;\\
J. Cao, W.-L. Yang, K. Shi and Y. Wang, {\it Nucl. Phys. } {\bf B 886} (2014), 185;\\
J. Cao, W.-L. Yang, K. Shi and Y. Wang, {\it Nucl. Phys. }{\bf B 877} (2013), 152.

\bibitem{Li14} Y.-Y. Li, J. Cao, W.-L. Yang, K. Shi, Y. Wang, {\it Nucl. Phys.} {\bf B 879}
(2014), 98.
\bibitem{Zha13} X. Zhang, J. Cao, W.-L. Yang, K. Shi, Y. Wang, {\it J. Stat. Mech.} (2014), P04031.
\bibitem{Hao14} K. Hao, J. Cao, G.\,-L. Li, W.\,-L. Yang, K. Shi and Y. Wang,
{\it JHEP} {\bf 06} (2014), 128.

\bibitem{Bas07} P. Baseilhac and K. Koizumi, {\it J. Stat.
Mech.\/} (2007), {\bf P09006}.

\bibitem{Skl92} E. K. Sklyanin, {\it Lect. Notes Phys.} {\bf 226} (1985), 196;\\
E. K. Sklyanin, {\it J. Sov. Math. } {\bf 31} (1985), 3417; \\
E. K. Sklyanin, {\it Prog. Theor. Phys. Suppl.} {\bf 118} (1995), 35.

\bibitem{Nic12} G. Niccoli, {\it Nucl. Phys.} {\bf B 870} (2013), 397; \\
G. Niccoli, {\it J. Phys.} {\bf  A 46} (2013), 075003.
\bibitem{Fad14} S. Faldella, N. Kitanine and G. Niccoli, {\it J. Stat. Mech.} (2014), P01011.
\bibitem{Kit14} N. Kitanine, J.-M. Maillet and G. Niccoli, {\tt arXiv:1401.4901}.

\bibitem{Fra08} H. Frahm, A. Seel and T. Wirth, {\it Nucl. Phys.} {\bf B 802} (2008), 351.

\bibitem{Veg93} H.\,J. de Vega and A. Gonz\'alez-Ruiz, {\it J. Phys.} {\bf A 26}, L519;\\
S. Ghoshal and A.\,B. Zamolodchikov, {\it Int. J.
Mod. Phys.\/} {\bf A 9} (1994), 3841.

\bibitem{Lor02} A. Di Lorenzo, L. Amico, K. Hikami, A. Osterloh, G. Giaquinta, {\it Nucl. Phys.}
{\bf B 644} (2002), 409.





\end{thebibliography}
\end{document}